\documentclass[epj]{webofc}
\usepackage[varg]{txfonts}   % Web of Conferences font
\wocname{EPJ Web of Conferences}
\woctitle{INPC 2013}
\begin{document}
\title{Emissivity and conductivity of parton-hadron matter}
%
% subtitle is optional

\author{O.~Linnyk\inst{1}\fnsep\thanks{\email{olena.linnyk@theo.physik.uni-giessen.de}}
\and    E.L.~Bratkovskaya\inst{2} \and
        W.~Cassing\inst{1} \and
        V.P.~Konchakovski\inst{1} \and
        V.~Ozvenchuk\inst{3}
}

\institute{ Institut f\"ur Theoretische Physik, %
  Universit\"at Giessen, %
%  Heinrich--Buff--Ring 16, %
  35392 Giessen, %
  Germany %
\and
Institut f\"ur Theoretische Physik, %
 Johann Wolfgang Goethe Universit\"at, %
%% Max-von-Laue-Str. 1, %
 60438 Frankfurt am Main, %
 Germany; %
Frankfurt Institute for Advanced Studies, %
 60438 Frankfurt am Main, %
 Germany;
\and
          SUBATECH, UMR 6457, Laboratoire de Physique Subatomique et des
Technologies Associ\'ees, %
University of Nantes - IN2P3/CNRS - Ecole des Mines de Nantes, %
% 4 rue Alfred Kastler, %
44072 Nantes Cedex 03, %
France;
          }

\abstract{%
  We investigate the properties of the QCD matter across the
  deconfinement phase transition. In the scope of the parton-hadron string
  dynamics (PHSD) transport approach, we study
  the strongly interacting matter in
  equilibrium as well as the out-of equilibrium dynamics of
  relativistic heavy-ion collisions. We present here in particular the
  results on the electromagnetic radiation, i.e. photon and dilepton
  production, in relativistic heavy-ion collisions and the relevant
  correlator in equilibrium, i.e. the
  electric conductivity. By comparing our calculations for the heavy-ion
collisions to the available data, we determine the relative
importance of the various production sources and address the
possible origin of the observed strong elliptic flow $v_2$ of direct
photons. }
\maketitle
%
%\section{Introduction}
%\label{intro}
\vspace{-0.2cm}

The electromagnetic emissivity of strongly interacting matter is a
subject of longstanding interest and is explored in particular in
relativistic nucleus-nucleus collisions, where the photons (and
dileptons) measured experimentally provide a time-integrated picture
of the collision dynamics.
 The recent observation by the PHENIX
Collaboration~\cite{PHENIX1} that the elliptic flow $v_2(p_T)$ of
'direct photons' produced in minimal bias Au+Au collisions at
$\sqrt{s_{NN}}=200$~GeV is comparable to that of the produced pions
was a surprise and in contrast to the theoretical expectations and
predictions~\cite{Chatterjee:2005de,Liu:2009kq,Dion:2011vd,Dion:2011pp,Chatterjee:2013naa}.
%Indeed, the photons produced by partonic interactions in the
%quark-gluon plasma phase have not been expected to show considerable
%flow because they are dominated by the emission in the initial phase
%before the elliptic flow fully develops.
% On the other hand, the
%dominant hadronic sources of photon production -- decays of mesons
%-- have been subtracted by the PHENIX Collaboration from the total
%photon spectrum using a model-independent method~\cite{PHENIX1} and
%therefore do not explain the observed strong momentum anisotropy of
%the direct photons.
%This has lead also to the suggestion that the
%photon $v_2$ observed might be a signature for more unconventional
%sources such as the pre-equilibrium gluon interaction with the
%magnetic field~\cite{Bzdak:2012fr,Basar:2012bp}, enhanced emission
%of photons at the QGP surface~\cite{Goloviznin:2012dy} or novel
%assumptions for the transverse parton acceleration in the
%QGP~\cite{Pantuev:2011yh,vanHees:2011vb}.
%
We have investigated the spectra and elliptic flow of dileptons and
photons from relativistic heavy-ion collisions in the scope of the
covariant transport approach Parton-Hadron String Dynamics (PHSD) in
Refs.~\cite{olena2010,Linnyk:2011hz,Linnyk:2011vx,Linnyk:2012pu,Linnyk:2013hta}.
The {PHSD} model~\cite{CasBrat,BrCa11} is an off-shell transport
approach that consistently describes the full evolution of a
relativistic heavy-ion collision from the initial hard scatterings
and string formation through the dynamical deconfinement phase
transition to the quark-gluon plasma as well as hadronization and
the subsequent interactions in the hadronic phase. The two-particle
correlations {resulting from the finite width of the parton spectral
functions} are taken into account dynamically {in the PHSD} by means
of the {generalized} off-shell transport equations~\cite{Cass_off1}
that go beyond the mean field or Boltzmann
approximation~\cite{Cassing:2008nn}. The transport theoretical
description of quarks and gluons in the PHSD is based on the
Dynamical Quasi-Particle Model (DQPM) for partons that is
{constructed} to reproduce lattice QCD (lQCD) results for the
entropy density, energy density and pressure as functions of
temperature for the quark-gluon plasma in thermodynamic equilibrium.
In the hadronic sector, PHSD is equivalent to the Hadron-String
Dynamics (HSD) approach~\cite{Cass99,Brat97,Bratkovskaya:2008iq}.
For details about the DQPM model and the off-shell transport
approach we refer the reader to the review
Ref.~\cite{Cassing:2008nn}.

\vspace{-0.2cm}
\section{Electric conductivity}
\label{sec-cond} \vspace{-0.2cm}

The photon emission rate from the QGP or the hadronic system is
controlled by the electric conductivity $\sigma_0$. The electric
conductivity of hot QCD matter at various temperatures $T$ was
studied within the PHSD approach for interacting partonic/hadronic
systems in a finite box with periodic boundary conditions in
Ref.~\cite{Cassing:2013iz} by computing the response of the
strongly-interacting system in equilibrium to an external electric
field. We found a sizeable temperature dependence of the ratio
$\sigma_0/T$.
The actual results for the ratio $\sigma_0/T$ versus the scaled
temperature $T/T_c$ are displayed in Fig.~\ref{v2_dir} (left hand
side) by the full round symbols. We observe a decreasing ratio
$\sigma_0/T$ with $T/T_c$ in the hadronic phase, a minimum close to
$T_c$ and an approximately linear rise  with $T/T_c$ above $T_c$
(=158 MeV). The conductivity in the partonic phase from PHSD can be
described by
\begin{equation} \label{e6} \frac{\sigma_0(T)}{T} \approx 0.01 +
0.16 \frac{T-T_c}{T_c}
\end{equation}
for $T_c\! \le \! T \! \le \! 2.2T_c$. The lQCD
results~\cite{l1,l2,l3,l4,l5} are represented in Fig.~1 (left hand
side) by symbols with error bars (using $C_{EM}=2 e^2/3$,
$e^2=4\pi\alpha$, $\alpha=1/137$). The linear rise with temperature
is supported by the recent lattice results from
Ref.~\cite{Amato:2013naa} (not explicitly presented in this plot).
In view of the pQCD prediction of a constant asymptotic value for
$\sigma_0/T \approx 5.9769/e^2 \approx65$ in leading order of the
coupling~\cite{l1,pQCD}, such a linear rise of the ratio with
temperature might be surprising, but it can be understood in simple
terms within the relaxation time approach with the DQPM interaction
width as demonstrated in Ref.~\cite{Cassing:2013iz}.
%
%The ratio drops in the hadronic phase with
%temperature
%$T$, shows a minimum close to $T_c$ and becomes approximately
%constant ($\sim$0.3) above $\sim\!5T_c$.
Our findings imply that the
QCD matter even at $T \! \approx \! T_c$ is a much better electric
conductor than $Cu$ or $Ag$ (at room temperature).

%\begin{figure}[h]
%\centering
%\includegraphics[width=0.45\textwidth,clip]{lQCDcomparison.eps}
%      \hspace{0.5cm}
%\includegraphics[width=0.49\textwidth,clip]{STAR_MB.eps}
%%
%\protect\caption{{\bf Left hand side: }Rates for dilepton production
%from the back-to-back $q+\bar q$ interactions in a thermal medium.
%PHSD results are shown by the red solid line, which is the sum of
%the $q+\bar q\to e^+e^-$ (blue dash line) and $q+\bar q\to g+e^+e^-$
%(green dash) contributions. The lattice QCD calculations are
%presented by the line with round symbols, the HTL results  by the
%magenta dash-dot line, the Gluon Condensate  rate by the orange
%dash-dot-dot line and leading order (Born) prediction  by the black
%dot line. {\bf Right hand side:} The PHSD results for the {invariant
%mass} spectra of inclusive {dileptons in Au+Au} collisions at
%$\sqrt{s{_{NN}}}$ = 200 GeV for 0 - 80 \% centrality within the cuts
%of the STAR experiment. }
%\end{figure}

\vspace{-0.2cm}
\section{Photon production in heavy-ion collisions}
\label{sec-1} \vspace{-0.2cm}

\begin{figure}
\centering
\includegraphics[width=0.53\textwidth,clip]{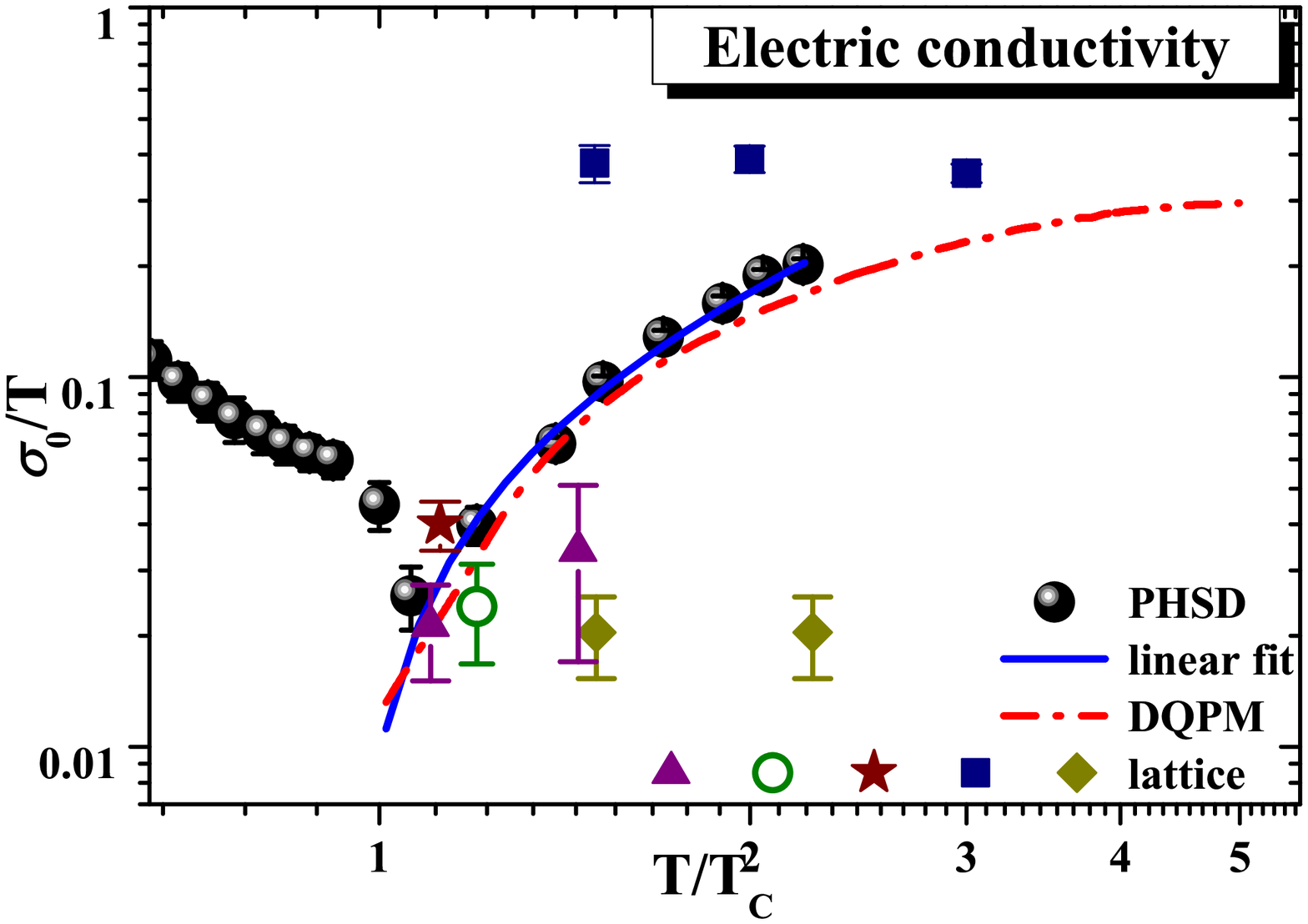}
      \hspace{0.05cm}
\includegraphics[width=0.43\textwidth,clip]{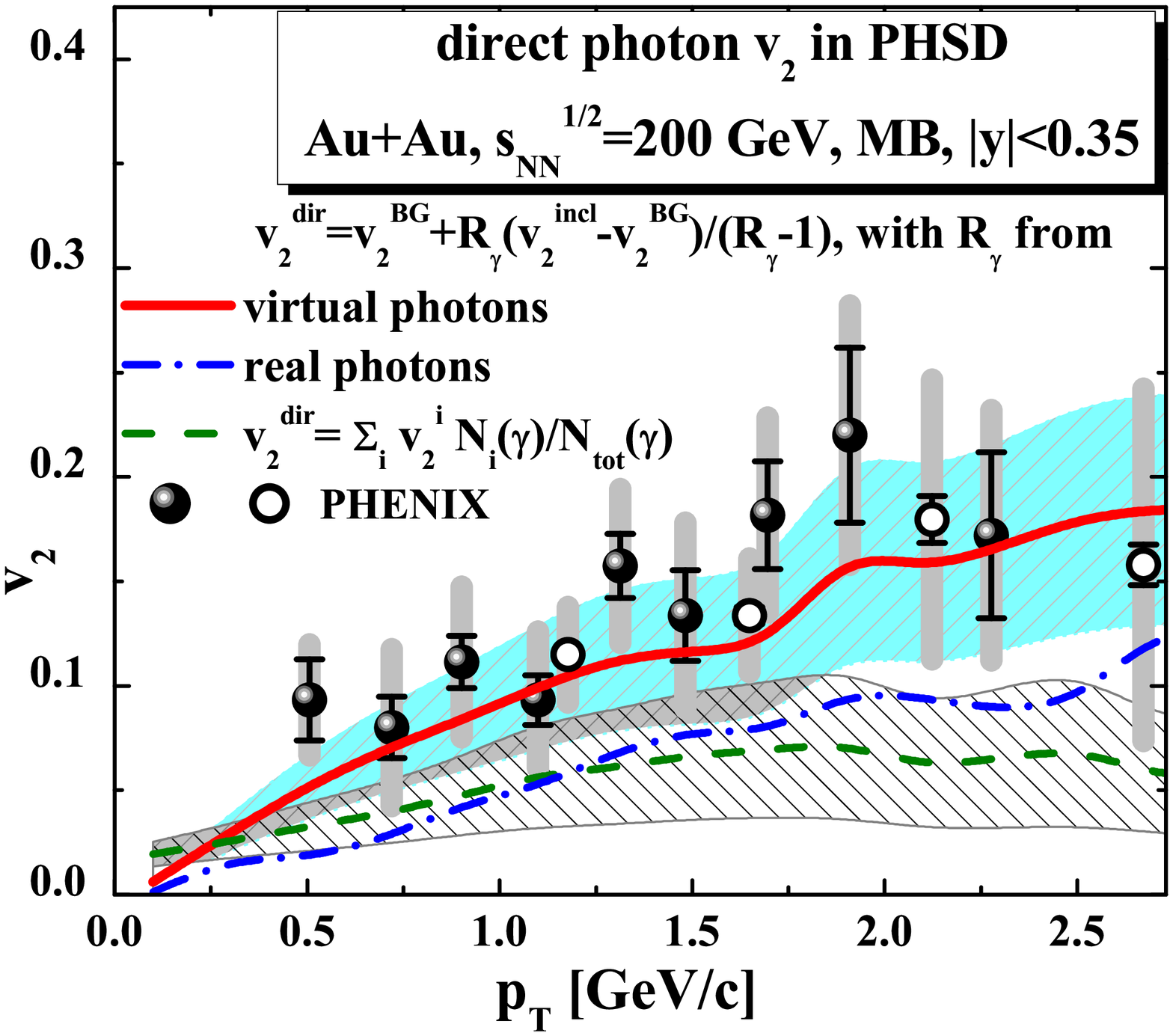}
\protect\caption{{\bf Left hand side:} The ratio of electric
conductivity to temperature $\sigma_0/T$ as a function of the scaled
temperature $T/T_c$ ($T_c$ = 158 MeV) for hot QCD matter. The full
round symbols show the PHSD results, the solid blue line is the
linear fit to the PHSD results (above $T_c$), while the dash-dotted
red line gives the corresponding ratio in the relaxation-time
approach (employing the DQPM parameters). The various symbols with
error bars represent the results from lattice calculations:
triangles -- quenched QCD results in the continuum limit with
Wilson-Clover fermions and renormalized vector currents from
Refs.~\cite{l1}, diamonds -- quenched QCD with staggered fermions
from Ref.~\cite{l2}, squares -- earlier work in the quenched QCD
with staggered fermions on smaller lattices from Ref.~\cite{l3},
star -- quenched SU(2) lattice gauge theory from Ref.~\cite{l4},
open circle -- QCD with two dynamical flavors of Wilson-Clover
fermions from Ref.~\cite{l5}.
 {\bf
Right hand side:} Elliptic flow of direct photons (hadron decays
excluded) in the PHSD approach for minimal bias Au+Au collisions at
$\sqrt{s_{NN}}=200$~GeV in comparison to the data from
Refs.~\cite{PHENIX1,Tserruya:2012jb}. The results from the PHSD are
displayed by the solid red line, equation (\protect\ref{dir1}), and
by the dashed green line, by applying Eq.~(\protect\ref{dir2}). }
\vspace{-0.6cm} \label{v2_dir}
\end{figure}

In  Ref~\cite{Linnyk:2013hta} we have applied the PHSD approach to
photon production in Au+Au collisions at $\sqrt{s_{NN}}=200$~GeV and
studied the transverse momentum spectrum and the elliptic flow $v_2$
of photons from hadronic and partonic production channels.
%The following sources for photon
%production were incorporated: the interactions of off-shell quarks
%and gluons in the strongly interacting quark-gluon plasma (sQGP)
%($q+\bar q\to g+\gamma$ and
% $q(\bar q)+g\to q(\bar q)+\gamma$), the decays of hadrons
%($\pi\to\gamma+\gamma$, $\eta\to\gamma+\gamma$,
%$\omega\to\pi+\gamma$, $\eta'\to\rho+\gamma$, $\phi\to\eta+\gamma$,
%$a_1\to\pi+\gamma$) as well as their interactions
%($\pi+\pi\to\rho+\gamma$, $\rho+\pi\to\pi+\gamma$, and meson-meson
%bremsstrahlung $m+m\to m+m+\gamma$).
%
We found that the PHSD calculations reproduce the transverse
momentum spectrum of direct photons as measured by the PHENIX
Collaboration in Refs.~\cite{PHENIXlast,Adare:2008ab}.
%Our
%microscopic calculations access the channel decomposition of the
%observed direct photon spectrum and show that the photons produced
%in the QGP constitute slightly less than 50\% with the rest being
%distributed among the other channels: mesonic interactions, decays
%of massive hadronic resonances and the initial hard scatterings.
%
Furthermore, the PHSD also describes the data on the elliptic flow
of inclusive photons~\cite{Linnyk:2013hta}. In order to extract the
flow of 'direct photons' from the inclusive one, the hadron decay
background has to be subtracted. This can be done by two procedures
which we describe below. %Firstly, the flow of the direct photon
%channels can be combined as a weighted sum (procedure 1). In this
%case, we take into account only the binary channels (the parton
%scatterings in the QGP as well as hadron reactions). The hadron
%decay photons does not enter the sum. The second possibility
%(procedure 2) is to follow the same background subtraction procedure
%as in the experiment (i.e. estimating the hadron decay contributions
%and subtracting their flow from the inclusive photon flow with
%relative weight).

{\bf \em Procedure 1.} We calculate the direct photon $v_2$ (in
PHSD) by summing up the elliptic flow of the individual channels
contributing to the direct photons, using their contributions to the
spectrum as the relative $p_T$-dependent weights, $w_i(p_T)$, i.e.
\begin{eqnarray} \label{dir2} &  v_2 (\gamma^{dir}) = \sum _i  v_2 (\gamma^{i})
w_i (p_T) =  {\sum _i  v_2 (\gamma^{i}) N_i (p_T)}/{\sum_i N_i
(p_T)}, &\\
&\mbox{where } i=(q\bar q \! \to \! g \gamma, q g \! \to \! q
\gamma, \pi \pi/\rho \! \to \! \rho/\pi \gamma, m m \! \to \! m m
\gamma, \mbox{pQCD}). & \nonumber \end{eqnarray}
The index $i$ denotes the binary production channels, both the
partonic quark-gluon interaction channels and the meson reactions
which cannot be separated presently experimentally by
model-independent methods.  The direct photon elliptic flow
calculated in this way is presented in Fig.~\ref{v2_dir} (right hand
side) by the dashed green line and clearly underestimates the PHENIX
data from Ref.~\cite{PHENIX1,Tserruya:2012jb} in line with
Refs~[2-6].

{\bf \em Procedure 2.} The experimental collaboration has extracted
the elliptic flow of direct photons $v_2(\gamma^{dir})$ from the
measured inclusive photon $v_2(\gamma^{incl})$ by subtracting the
hadron decay sources ($\pi_0$, $\eta$, $\omega$, $\eta'$, $\phi$,
$a_1$) as follows~\cite{PHENIX1}:
\begin{eqnarray} \label{dir1}
& v_2 (\gamma^{dir}) =  \left(R_\gamma v_2(\gamma^{incl}) - v_2
(\gamma^{BG})\right)/\left(R_\gamma -1\right) = v_2 (\gamma^{BG}) +
\frac{R_\gamma }{R_\gamma -1} ( v_2(\gamma^{incl})-v_2(\gamma^{BG}))
&
\end{eqnarray}
where $$R_\gamma=N^{incl}/N^{BG}$$ denotes the ratio of the
inclusive photon yield to that of the ``background" (i.e. the
photons stemming from the decays of $\pi_0$, $\eta$, $\omega$,
$\eta'$, $\phi$ and $a_1$ mesons), $v_2(\gamma^{BG})$ is the
elliptic flow of the background photons; one can assume
$v_2(\gamma^{BG})\approx v_2(\pi_0)$~\cite{Linnyk:2013hta}.
The ratio $R_\gamma$ was obtained experimentally in
Ref.~\cite{Adare:2008ab} by analyzing the yield of dileptons with
high transverse momentum $p_T>1$~GeV and low invariant mass $M$, the
$R_\gamma$ values obtained in Ref.~\cite{Adare:2008ab} for the
photon transverse momenta $1<p_T<3$~GeV are on average $1.2\pm0.3$.
We recall here that we have studied the dilepton production at the
top RHIC energy within the PHSD approach in
Ref.~\cite{Linnyk:2011vx}. The PHSD results reproduce well the
PHENIX and STAR dilepton data differentially in the invariant mass
$M$ and transverse momentum $p_T$, only underestimating the excess
observed by PHENIX at low $M$ and low $p_T$. Note, however, that for
the relatively high transverse momenta of dileptons ($p_T>1$~GeV)
the agreement of the PHSD calculations with the PHENIX data is quite
good. We obtained $R_\gamma\approx1.2$ by analyzing the yield of
dileptons in the invariant mass window
$M=0.15-0.3$~GeV~\cite{Linnyk:2013hta}.
Alternatively, we can use the calculated inclusive {\em real} photon
spectrum to find the ratio
$R_\gamma=N(\gamma)^{incl}/N(\gamma)^{BG}$. In this case we obtained
$R_\gamma\approx1.08$ from the real photons in
PHSD~\cite{Linnyk:2013hta}. The difference between the values of
$R_\gamma$ -- extracted from the dilepton spectra and the real
photon spectra -- is caused by the fact that in the dilepton mass
window $M=0.15-0.3$~GeV the background from the pion decays
effectively ``dies out", while the pion decay contribution is
prominent for $M\to0$.

Following the procedure of equation (\ref{dir1}) in the PHSD, we
obtain the red solid line in Fig.~\ref{v2_dir} (right hand side), if
we use the ratio $R_\gamma$ from the virtual photons in the
invariant mass window $M=0.15-0.3$~GeV, and the blue dash-dotted
line, if we use the ratio $R_\gamma$ from the calculated real photon
spectrum. The two lines differ by about a factor of two. The
difference between the two extraction procedures for the direct
photon flow $v_2(p_T)$ can be attributed to different definitions
for the ratio of the inclusive and background photons ($R_\gamma$).

%\section{Conclusions}
In conclusion,
%our calculations show that the photon production in
%the QGP is dominated by the early phase (similar to hydrodynamic
%models) and is localized in the center of the fireball, where the
%collective flow is still rather low, i.e. on the 2-3 \% level, only.
%
%Thus, the strong $v_2$ of direct photons - which is comparable to
%the hadronic $v_2$ - in PHSD is attributed to hadronic channels,
%i.e. to meson binary reactions which are not subtracted in the data.
%On the other hand, the strong $v_2$ of the 'parent' hadrons, in
%turn, stems from the interactions in the QGP via collisions and the
%partonic mean-filed potentials. Our
our findings imply that there is presently no clear signal for
'unconventional physics' (beyond the strong interaction on the
partonic and hadronic level) in the photon data from the PHENIX
Collaboration within error bars. The strong $v_2$ of direct photons
- which is comparable to the hadronic $v_2$ - in PHSD is attributed
to hadronic channels, i.e. to meson binary reactions which are not
subtracted in the data. On the other hand, the strong $v_2$ of the
'parent' hadrons, in turn, stems from the interactions in the QGP.
%via collisions and the
%partonic mean-filed potentials.
Accordingly, the presence of the QGP shows up 'indirectly' in the
direct photon elliptic flow.

%
% BibTeX or Biber users please use (the style is already called in the class, ensure that the "woc.bst" style is in your local directory)
%\bibliography{PHSDdilept}
%\begin{thebibliography}{}
%Journal Author, Journal \textbf{Volume}, page numbers (year)
%Book Author, \textit{Book title} (Publisher, place, year) page numbers
% etc
%\end{thebibliography}

\end{document}